# A Pearson- Dirichlet random walk

Gérard Le Caër

*Institut de Physique de Rennes, UMR UR1-CNRS 6251, Université de Rennes I, Campus de Beaulieu, Bâtiment 11A, F-35042 Rennes Cedex, France*

Tel :+33 2 23 23 56 28

Fax :+33 2 23 23 67 17

*E-mail address*:

gerard.le-caer@univ-rennes1.fr

Keywords : Pearson-Rayleigh random walks, simplex, hypersphere, uniform distribution, Dirichlet distribution, Liouville distribution, division of the unit interval

## Abstract

A constrained diffusive random walk of $n$ steps in $\mathbb{R}^d$ and a random flight in $\mathbb{R}^d$, which can be expressed in the same terms, were investigated independently in recent papers (J. Stat. Phys. 127, 813 (2007), J. Theor. Probab. 20, 769 (2007) and J. Stat. Phys. 131, 1039 (2008)). The $n$ steps of the walk are identically and independently distributed random vectors of exponential length and uniform orientation. Conditioned on the sum of their lengths being equal to a given value $l$, closed-form expressions for the distribution of the endpoint of the walk were obtained altogether for any $n$ for $d=1, 2, 4$. Uniform distributions of the endpoint inside a ball of radius $l$ were evidenced for a walk of three steps in 2D and of two steps in 4D.

The previous walk is generalized by considering step lengths which are distributed over the unit $(n-1)$ simplex according to a Dirichlet distribution whose parameters are all equal to $q$, a given positive value. The walk and the flight above correspond to $q=1$. For any space dimension $d \geq 3$, there exist, for integer and half-integer values of $q=q(d)$, two families of "Pearson-Dirichlet" walks and only two which share a common property. For any $n$, the $d$ components of the endpoint are jointly distributed as are the $d$ components of a vector uniformly distributed over the surface of a hypersphere of radius $l$ in a space $\mathbb{R}^k$ whose dimension $k$ is an affine function of $n$ for a given $d$. Five additional walks, with a uniform distribution of the endpoint in the inside of a ball, are found from known finite integrals of products of powers and Bessel functions of the first kind. They include four different walks in $\mathbb{R}^3$, two of two steps and two of three steps, and one walk of two steps in $\mathbb{R}^4$. Pearson-Liouville random walks, obtained by distributing the total lengths of the previous Pearson-Dirichlet walks according to some specified probability law, are finally discussed.

## 1. Introduction

Considering a particle moving in a random environment and undergoing elastic collisions at uniformly distributed point obstacles, Franceschetti [1] defined a variant of the Pearson-Rayleigh random walk in $\mathbb{R}^d$ [2]: the $n = m+1$ steps of the walk are independent and identically distributed (i.i.d.) $d$-dimensional random vectors whose lengths have an exponential distribution and whose orientations are uniform. Constraining the total travelled length to be equal to a given $l(>0)$, Franceschetti derived in 1D and in 2D the conditional probability density function (pdf) $p_{d,n}^{(l)}(\boldsymbol{r})$ of the endpoint $\boldsymbol{r} = \boldsymbol{r}_n^{(d)}$ of the walk. The latter density depends only on the distance $r = \|\boldsymbol{r}\|$ from the endpoint to the origin as the walk is statistically invariant by any orthogonal transformation.

For a 2D space, the pdf $p_{d,n}^{(l)}(\boldsymbol{r})$ reads [1]:

$$p_{2,n}^{(l)}(\boldsymbol{r}) = \frac{(n-1)}{2\pi l^2}\left(1-\frac{r^2}{l^2}\right)^{\frac{n-3}{2}} \quad (r<l) \; n = 2,3,... \qquad (1)$$

Franceschetti concluded that a walker is more likely to end its walk near the boundary of the disc of radius $l$ when making less than three steps in 2D and near the origin when making more than three steps. By making exactly three steps, the endpoint is uniformly distributed inside the disc of radius $l$ (eq. 1). The value $n = 3$ in 2D was considered as a 'critical transition point' in the behaviour of the random walk. A uniform distribution was similarly concluded to occur for two steps in 1D. Indeed, a continuous and uniform density exists in that case between $-l$ and $+l$ but a delta peak with a weight of 1/4 is further located at each boundary of the interval $[-l, l]$. The actual distribution is then a mixture of a discrete and of a uniform distribution with equal weights. In any case, the question naturally arose as to whether it is possible to find a uniform distribution for another couple $(d, n)$. As calculations were considered to become intractable in dimensions higher than $d = 2$, Franceschetti [1] derived a necessary condition in the form of a relation between $n$ and $d$, $d(n-1) = 4$, from the calculation of the second moment of the distance of the endpoint to the origin. That relation is only satisfied by the two couples $(d = 2, n = 3)$ and $(d = 4, n = 2)$. The previous walk was formulated in terms of scattering of particles by García-Pelayo [3]. He concluded that the $(d=2,n=3)$ walk is the sole walk of the whole family whose endpoint distribution is uniform.

As shown in section 2, the aforementioned walk is directly related to a random flight performed by a particle in $\mathbb{R}^d$ which starts from the origin at time $t=0$, moves with a constant and finite velocity $c$ in an initial random direction uniformly distributed on the unit hypersphere. It flies until it chooses a new direction at a random time determined by a homogeneous Poisson process, independently of the previous direction. The particle moves with velocity $c$ until the next Poisson signal obliges it to change its direction again and so on [4-5]. The conditional pdf of the position of the particle at time $t$, given the number $m$ of Poisson events that occurred up to $t$, was obtained for any $m$ for $d=2$ and for $d=4$ [4-5]. For a given time interval $t$, the total length of the flight is then fixed. Replacing $ct$ by $l$ in the corresponding pdf's, the density of eq. 1 is obtained for $d=2$ while, for $d=4$, [4-5]:

$$p_{4,n}^{(l)}(\boldsymbol{r})=\frac{n(n-1)}{\pi^2 l^4}\left(1-\frac{r^2}{l^2}\right)^{n-2} \quad (r<l) \quad n=2,3,... \tag{2}$$

That distribution is identical with the one found in the present work for a walk of $n$ steps in 4D (sections 3 and 4). A uniform distribution exists then for $n=2$ in 4D [4-5] in contradiction with the conclusion of García-Pelayo [3]. The latter is explained by an error in equation 7 of [3] in which $4!/6!$ must be replaced by $(2+s)!/(4+s)!$. A closed-form expression was further obtained by Kolesnik [5] for a walk of two steps in 3D, $p_{3,2}^{(l)}(\boldsymbol{r})=\frac{1}{4\pi l^2 r}\ln\left(\frac{1+r/l}{1-r/l}\right)$ $(r<l)$. Without loss of generality, we take hereafter the walk length $l$ as equal to 1 and we drop the “$l$“ from the previous notations.

The existence of uniform distributions stems actually from a more general property of the $d$-variate pdf $p_{d,n}(\boldsymbol{r})$ for $d=2,4$. The latter is indeed that of $d$ components of a $k$-dimensional unit vector, $\boldsymbol{u}^{(k)}$, whose tip spans uniformly the surface of a hypersphere in $\mathbb{R}^k$, where $k(\geq d)$ is an affine function of $n$ for a fixed $d$ [5]( $k=n+1$ in 2D and $k=2n+2$ in 4D). In other words, it suffices to project $\boldsymbol{u}^{(k)}$ in $\mathbb{R}^d$ to get the sought-after pdf of $\boldsymbol{r}_n^{(d)}$. From now on, we will name for brevity "hyperspherical uniform" (HU) a walk whose endpoint distribution exhibits the latter property. In that case, the line of reasoning based on a recurrence relation obeyed by the associated characteristic functions becomes natural [5]. We realized independently of [5] that the pdf given by eq. 1 exhibits the HU property (emails were exchanged on that topic with M. Franceschetti (6-7 June 2007)). That observation motivated the present work.

The step lengths of the previous random walk and random flight are uniformly distributed over the unit $(n-1)$ simplex. This distribution is a Dirichlet distribution whose parameters are all equal to 1 (section 2). More generally, we will consider "Pearson-Dirichlet" random walks of $n$ steps in $\mathbb{R}^d$ denoted by $PD(d,n,q)$. The $n$ steps of a walk $PD(d,n,q)$ are i.i.d. random vectors in $\mathbb{R}^d$ whose orientations are uniform and whose lengths are distributed over the unit $(n-1)$ simplex according to a Dirichlet distribution whose parameters are chosen to be all equal to a given positive value $q$ (figure 1). The latter $n$-variate distribution of the step length is invariant by any permutation. A simple recurrence relation, similar to that derived for $q=1$ [5], will be found to result from the latter choice. Among the variants of the Pearson walk with unequal step sizes, a walk with shrinking step lengths was investigated very recently in 2D (the length of the $i$th step is $\lambda^{i-1}(\lambda<1)$) [6]. Interestingly, the existence of a critical value of $\lambda$ is evidenced which marks a change of behaviour of the distribution of the distance of the endpoint to the origin.

The aim of the present work is to find all hyperspherical uniform walks associated with the unit hypersphere of $\mathbb{R}^k$ for a walk of $n$ steps in $\mathbb{R}^d$, denoted hereafter as $HU(d,n,q,k)$, among the Pearson-Dirichlet walks $PD(d,n,q)$. The pdf's of the endpoints of such HU walks, denoted, $p_{d,n,q}(\boldsymbol{r})$, will then be readily obtained from the hypersphere space dimension $k$ and the possible occurrence of other uniform distributions will be examined.

The characteristics of the Dirichlet distribution will be first briefly recalled. A necessary condition for a walk to be a HU walk will then be established and a recurrence equation between the characteristic functions associated with walks of $n-1$ and $n$ steps will be solved to establish that the selected walks are indeed HU. Besides the two uniform walks in 2D and in 4D reported for $q=1$ [1,4-5], five additional uniform walks, four in 3D and one in 4D, will be shown to be associated with known finite integrals of products of powers and Bessel functions of the first kind.

A Pearson-Liouville random walk will be finally defined from the previous HU walks by letting the total walk length $l$ vary according to some pdf $f(l)$. Illustrative examples of random walks of $n$ i.i.d. steps in $\mathbb{R}^d$, whose lengths have a gamma distribution with a shape parameter either equal to $d-1$ $(d\geq 2)$ or to $\frac{d}{2}-1$ $(d\geq 3)$, will be considered.

## 2. The uniform distribution on the unit m-simplex and the Dirichlet distribution

The unit $(n-1)$ simplex $\boldsymbol{S}_{n-1}$ is the set defined by:

$$\boldsymbol{S}_{n-1}=\left\{(l_1,l_2,..,l_n)\in\mathbb{R}^n:\sum_{i=1}^{n}l_i=1 \text{ and } l_i\geq 0 \text{ for any } i\right\} \tag{3}$$

The joint pdf $p_e(s_1,s_2,..,s_n)$ of i.i.d. exponential step lengths $s_i>0(i=1,..,n)$, with a scale parameter of 1, is simply given by:

$$p_e(s_1,s_2,..,s_n)=\exp\left(-\sum_{i=1}^{n}s_i\right) \tag{4}$$

Defining first their sum as $s=\sum_{i=1}^{n}s_i$ and $l_i=s_i/s$ $(i=1,..,n)$, then changing the set of variables from $(s_1,s_2,..,s_n)$ to $(l_1,l_2,..l_m,s)$, the following joint pdf is simply obtained from the Jacobian of the transformation which is equal to $s^m$ :

$$p_u(l_1,l_2,..l_m,s)=m!\left(\frac{1}{m!}s^m\exp(-s)\right)=p(l_1,l_2,..l_m)\times p_S(s) \tag{5}$$

Then $\boldsymbol{l}^{(m)}=(l_1,l_2,..l_m)$ and $s$ are independent. As expected the distribution of $s$, which is the sum of $n=m+1$ i.i.d. exponential random variables, is a gamma distribution with a shape factor of $n$. Further, the pdf :

$$p(l_1, l_2, .., l_m) = m! \quad (6)$$

is constant. Rescaling the total length to make $l = \sum_{k=1}^{n} l_k = 1$, the distribution of the step length of the walk defined by Franceschetti [1] is, by construction, uniform over the unit $(n-1)$ simplex $\boldsymbol{S}_{n-1}$. The same conclusion holds for the random flight considered in [4-5]. Indeed, the joint pdf of the times of occurrence of events from an homogeneous Poisson process in the time interval $(0,t]$, given that the number of events is $N(t) = m$, is $p\left(t_1, t_2, .., t_m \middle| N(t) = m\right) = \frac{m!}{t^m}$ $\left(0 < t_1 < t_2 < ... < t_m \le t\right)$ (see for instance [7] p. 277). The distribution of the inter-arrival times, in a time interval scaled down to unit length, is thus given by eq. 6 as the Jacobian of the transformation is 1. The latter distribution ($t = 1$) is too that of $m$ i.i.d. random variables $U_i\ (i = 1, .., m)$ which are uniformly distributed on the unit interval (0-1) [7-8]. When their values are sorted in increasing order, the ordered arrival "times" are denoted as $U_{(i)}\ (i = 1, .., m)$. Then, the multivariate pdf of the inter-arrival times, $V_{(i)} = U_{(i)} - U_{(i-1)}\ \left(i = 1, .., m\ ;\ U_{(0)} = 0\right)$, is uniform over the unit $(n-1)$ simplex [8]. The distribution of the endpoint of the walk studied by Franceschetti [1] and the conditional distribution of the particle after a random flight of duration $t$ investigated by Orsingher and DeGregorio [4] and by Kolesnik [5] are concluded to be identical after a replacement of $ct$ by $l$ (eqs 1 and 2). For a total length of 1, the step lengths have a common distribution which is a particular case of a Dirichlet distribution with all its parameters equal to 1.

The Dirichlet distribution is of common use in simplices. It is applied for instance in ecology, to model fragmentation or compositional data [9-10]. The walk performed by a donkey inside a tetrahedron, as constructed by Letac [11], leads to a stationary distribution which is Dirichlet. The Dirichlet distribution can be simply defined as follows [12]: consider a set of $n = m + 1$ independent gamma distributed random variables, $s_i\ (> 0,\ i = 1, .., n)$ $p(s_i) = s_i^{\alpha_i - 1} \exp(-s_i) / \Gamma(\alpha_i)$, where $\Gamma(x)$ is the Euler gamma function, with shape parameters $\alpha_i > 0\ (i = 1, .., n)$ and scale parameters of 1 (for simplicity) and define $l_j = s_j \Big/ \sum_{i=1}^{n} s_i \ \ (j=1,..,n)$. The distribution of $\boldsymbol{l}^{(m)}$ is a Dirichlet distribution $D_m\left(\boldsymbol{\alpha}^{(n)}\right)$ with parameters $\boldsymbol{\alpha}^{(n)} = (\alpha_1, ..., \alpha_m, \alpha_n)$. Using a method identical with the one used above, its pdf is established to be ([12], p. 17):

$$\begin{cases} p_m(l_1,..,l_m) = K\left(\boldsymbol{\alpha}^{(n)}\right)\prod_{i=1}^{n} l_i^{\alpha_i - 1} \\ l_n = 1 - \sum_{i=1}^{m} l_i, \quad l_i > 0, \ i = 1,..,n \end{cases} \tag{7}$$

where $K\left(\boldsymbol{\alpha}^{(n)}\right) = \Gamma(\alpha) \Big/ \left(\prod_{i=1}^{n}\Gamma(\alpha_i)\right)$, $\alpha = \sum_{i=1}^{n}\alpha_i$. Defining the vector $\boldsymbol{\beta}^{(n)} = (\beta_1, ..., \beta_m, 0)$ and $\beta = \sum_{i=1}^{n}\beta_i$, the moment $M_{\boldsymbol{\beta}} = \left\langle \prod_{i=1}^{m} l_i^{\beta_i} \right\rangle$ is simply obtained by noticing that it is related to the normalization constant of the Dirichlet distribution $D_m\left(\boldsymbol{\alpha}^{(n)} + \boldsymbol{\beta}^{(n)}\right)$, namely:

$$M_{\boldsymbol{\beta}} = \left\langle \prod_{i=1}^{m} l_i^{\beta_i} \right\rangle = K\left(\boldsymbol{\alpha}^{(n)}\right) \Big/ K\left(\boldsymbol{\alpha}^{(n)} + \boldsymbol{\beta}^{(n)}\right) = \left(\prod_{i=1}^{m} (\alpha_i)_{\beta_i}\right) \Big/ (\alpha)_{\beta} \tag{8}$$

where $(a)_r = \Gamma(a+r)/\Gamma(a)$ reduces to an ascending factorial, $(a)_r = a(a+1)..(a+r-1)$, when $r$ is an integer. The Dirichlet distribution has a notable amalgamation property [12]. If the $n$ components $l_1,..,l_n \left(\sum_{i=1}^{n} l_i = 1\right)$ of a vector, whose distribution is $D_m\left(\boldsymbol{\alpha}^{(n)}\right)$, are grouped into $k$ components $v_1,..,v_k \left(\sum_{i=1}^{k} v_i = 1\right)$, then the distribution of $(v_1,..,v_{k-1})$ is $D_{k-1}\left(\boldsymbol{\alpha}^{*(k)}\right)$ where each $\alpha_i^* \ (i=1,..,k)$ is the sum of the $\alpha_j$ 's corresponding to the components of the initial vector which add up to $v_i$. The marginal distribution of any component $l_i$ is then obtained by grouping the remaining components into a single one. Any component $l_i$ has a beta distribution [13], $beta(\alpha_i, \alpha - \alpha_i)$ with a pdf: $p_i(l_i) = \frac{\Gamma(\alpha)}{\Gamma(\alpha_i)\Gamma(\alpha - \alpha_i)} \times l_i^{\alpha_i - 1}(1 - l_i)^{\alpha - \alpha_i - 1}$ $(0<l_i<1)$. The amalgamation property results directly from the well-known fact that a sum of independent gamma random variables, with identical scale parameters and a priori different shape parameters, is still a gamma random variable with the same scale parameter and a shape parameter which is the sum of all shape parameters [7,12-13].

The method based on the generation of i.i.d. gamma random variables is the simplest method for simulating Dirichlet distributions on simplices ([14] for the case $\boldsymbol{\alpha}^{(n)} = (1,..,1)$) that was used in the present work for Monte-Carlo simulations of $PD(d,n,q)$. Finally, it is readily seen from the definition of the Dirichlet distribution that the conditional distribution of $l_j^* = \frac{l_{j+1}}{1-l_1}$, $(j = 1,..,m-1)$ , given $l_1$, is still a Dirichlet distribution (theorem 1.6 of Fang et al. [12]), $D_{m\text{-}1}\left(\boldsymbol{\alpha}^{(n\text{-}1)} = \left(\alpha_2,..,\alpha_{m,}\alpha_n\right)\right)$, with a pdf:

$$p_{m-1}\left(l_1^*,..,l_{m-1}^* | l_1\right) = p_{m-1}\left(l_1^*,..,l_{m-1}^*\right) = K\left(\boldsymbol{\alpha}^{(n\text{-}1)}\right)\left[\prod_{i=1}^{m} l_i^{*\alpha_{i+1}-1}\right] \tag{9}$$

As the distribution of $\boldsymbol{l}^{*(m-1)}$ is independent of the distribution of $l_1$, then $l_1$ is said to be neutral [15]. That property serves as a basis for a generalization of Dirichlet distributions ([16] and references therein). The step lengths of the walks considered in what follows, $(l_1, l_2, ..l_m)$, have almost all a Dirichlet distribution $D_m\left(\boldsymbol{q}^{(n)}\right)$, where $\boldsymbol{q}^{(n)}$ is the $n$-dimensional vector whose components are all equal to $q$ $(q > 0)$. In that case, the distribution $p_{m-1}\left(l_1^*,..,l_{m-1}^*\right)$ is simply $D_{m\text{-}1}\left(\boldsymbol{q}^{(n-1)}\right)$.

## 3. A necessary condition for a walk $PD(d,n,q)$ to be "hyperspherical uniform"

The general problem of obtaining closed-form expressions of the probability density of the endpoint of a walk of $n$ steps $PD(d,n,q)$ in $\mathbb{R}^d$ is intractable for any $n$ and $d$ . Rather than coping with an insoluble problem, we chose to search under the lamppost to find all Pearson-Dirichlet walks $PD(d,n,q)$ whose distributions of the endpoint $p_{d,n,q}(\boldsymbol{r})$ are simple to calculate. The conjunction of known finite integrals of products of two Bessel functions of the first kind with powers and the possible existence of recurrence relations for values of $q > 0$ other than 1 led us to select the hyperspherical uniform property as our lamppost.

The Dirichlet distribution of the vector $\boldsymbol{l}^{(m)}$ has a multivariate pdf given by:

$$p_m(l_1,..,l_m) = \frac{\Gamma(nq)}{\Gamma(q)^n} \times \left[\prod_{i=1}^{n} l_i^{q-1}\right] \tag{10}$$

with $l_n = 1 - \sum_{i=1}^{m} l_i, \quad l_i > 0, \; (i = 1,..,n)$ . The endpoint of a Pearson-Dirichlet walk $PD(d,n,q)$ is a vector of $\mathbb{R}^d$ which reads $(n \geq 2)$:

$$\boldsymbol{r}_n^{(d)} = \sum_{i=1}^{n} l_i \boldsymbol{u}_i^{(d)} \tag{11}$$

where the $\boldsymbol{u}_i^{(d)} = (u_i(1),..,u_i(d)),(i = 1,..,n)$, are $n$ independent unit vectors uniformly distributed over the surface of the hypersphere in $\mathbb{R}^d$. A simple necessary condition for a walk to be HU is that the even moments of a single component $r_1 = r_n(1)$ of $\boldsymbol{r}_n^{(d)}$ are equal to the even moments of any component of a unit vector uniformly distributed over the surface of a hypersphere in some space $\mathbb{R}^k$, where $k$ has to be determined. That necessary condition will thus provide all possible sets $(d,n,q,k)$ for which the sought-after property might hold. Actually, the moments $\langle r_1^2 \rangle$ and $\langle r_1^4 \rangle$ happen to suffice. The moment $\langle r_1^2 \rangle$ is just $n$ times the product of $\langle l_i^2 \rangle = \dfrac{q(q+1)}{(nq)(nq+1)}$ with $\langle u_i(j)^2 \rangle = \dfrac{1}{d}$, the cross-products being zero because the $\boldsymbol{u}_i^{(d)}$'s are independent and have zero means. Similarly, the moment $\langle r_1^4 \rangle$ is given by the sum: $n\langle l_1^4 \rangle \langle u_1^4(1) \rangle + 3n(n-1)\langle l_1^2 l_2^2 \rangle \langle u_1^2(1) \rangle^2$. These moments are (eqs 8 and A-3):

$$\begin{cases} \langle r_1^2 \rangle = \dfrac{1}{k} = \dfrac{q+1}{d(nq+1)} \\ \langle r_1^4 \rangle = \dfrac{3}{k(k+2)} = \dfrac{3(q+1)(q+2)(q+3)}{d(d+2)(nq+1)(nq+2)(nq+3)} + \dfrac{3(n-1)q(q+1)^2}{d^2(nq+1)(nq+2)(nq+3)} \end{cases} \tag{12}$$

We notice first that eq. 12 yields the expected result for $n=1$, namely $k=d$ as the endpoint of a walk of one step is by definition uniformly distributed over the surface of the hypersphere in $\mathbb{R}^d$. In the following, we take $n>1$. The moment $\langle r_1^2 \rangle$ gives:

$$k=\frac{d(nq+1)}{q+1} \tag{13}$$

which, when plugged into $\langle r_1^4 \rangle$ (eq. 15), should give a relation between $d,n$ and $q$. That relation simplifies actually to:

$$d^2 - 3(q+1)d + 2(q+1)^2 = 0 \tag{14}$$

which is independent of $n$. The two solutions, whose correctness is readily verified from eq. 12, and the corresponding "hyperspace" dimensions are:

$$\begin{cases} d' = q+1 & k' = nq+1 \\ d'' = 2(q+1) & k'' = 2(nq+1) \end{cases} \tag{15}$$

The previous necessary condition indicates thus that there are two possibilities: either $q(\geq 1/2)$ is an integer or it is a half-integer. When $q$ is an integer, at most two walks, might be HU, namely $HU(q+1,n,q,nq+1)$ and $HU(2q+2,n,q,2nq+2)$ for any number of steps, where $HU(d,n,q,k)$ is recalled to be associated with the unit hypersphere of $\mathbb{R}^k$ for a walk of $n$ steps in $\mathbb{R}^d$. A third possible HU walk in $\mathbb{R}^{2p+3}$ is found to be $HU(2p+3,n,q,n(2p+1)+2)$ for $q=(2p+1)/2$ .

Using a recurrence relation between the characteristic functions of the probability distributions of the endpoints of walks of $n-1$ and of $n$ steps, quite similar to that derived by Kolesnik [5] for $q=1$, we prove in the next section that the three previous walks are indeed hyperspherical uniform walks for any $n$. The necessary conditions of the present section will thus be found to be sufficient. Table 1 gathers the parameters needed to obtain the endpoint pdf $p_{d,n,q}(\boldsymbol{r})$ from eq. A-2 and consequently that of the distance of the endpoint to the origin $P_{d,n,q}(r)$ for all hyperspherical uniform walks of the $PD(d,n,q)$ family. The distribution of the square of the distance, $s = r^2$, is $beta(d/2, \delta+1)$. Once the latter distributions are known for $l = 1$, one gets immediately that:

$$\begin{cases} p_{d,n,q}^{(l)}(\boldsymbol{r}) = \dfrac{1}{l^d}\, p_{d,n,q}\left(\dfrac{\boldsymbol{r}}{l}\right) \\ P_{d,n,q}^{(l)}(r) = \dfrac{1}{l^d}\, P_{d,n,q}\left(\dfrac{r}{l}\right) \end{cases} \tag{16}$$

for any $l$. The pdf's given by eqs 1 and 2 [1, 4-5] are obtained for $q = 1$ from the second line of table 1 for a walk in 2D and from the third line for a walk in 4D. Table 2 collects all the characteristics of the uniform walks obtained from $p_{d,n,q}(\boldsymbol{r}) \propto (1-r^2)^{\delta}$ with $\delta = 0$ (table 1), and those of three additional walks derived in section 5.

**4. A recurrence relation**

To establish that a walk $PD(d,n,q)$ is hyperspherical uniform, $HU(d,n,q,k)$, it suffices to prove that the characteristic function (c.f.) of the probability distribution of the endpoint $\boldsymbol{r}_n^{(d)}$, or of a single component of it, is $\Omega_k(\rho)$. The latter c.f. is that of a unit vector whose tip is uniformly distributed over the surface of the hypersphere in $\mathbb{R}^k$ (eq. A-1 and appendix). The possible sets of values of $k$ are given in table 1 as a function of $d,n,q$. The conditional pdf $p_{m-1}(l_1^*, l_2^*, ..l_{m-1}^* | l_1)$, which is $D_{m\text{-}1}(\boldsymbol{q}^{(n\text{-}1)})$ ( eq. 9), allows us to express the endpoint of the walk of $n \geq 2$ steps in $\mathbb{R}^d$ as follows:

$$\boldsymbol{r}_n^{(d)} = l_1\boldsymbol{u}_1^{(d)} + \left[\sum_{i=2}^{n} l_i\boldsymbol{u}_i^{(d)}\right] = l_1\boldsymbol{u}_1^{(d)} + (1\text{-}l_1)\,\boldsymbol{r}_{n\text{-}1}^{(d)} \tag{17}$$

From eq. 17 and the marginal pdf, $p_1\left(l_1\right)=\frac{\Gamma\left(nq\right)}{\Gamma\left(q\right)\Gamma\left(\left(n-1\right)q\right)}l_1^{q-1}\left(1-l_1\right)^{(n-1)q-1}$, which results from the amalgamation property of the Dirichlet distribution (section 2), we obtain the characteristic function of the probability distribution of $\boldsymbol{r}_n^{(d)}$, $\Phi_{d,n,\boldsymbol{q}^{(n)}}\left(\rho\right)=\left\langle\exp\left(i\boldsymbol{\rho}.\boldsymbol{r}_n^{(d)}\right)\right\rangle$ $\left(\rho=\left\|\boldsymbol{\rho}\right\|\right)$:

$$\begin{cases}\Phi_{d,1}\left(\rho\right)=\Omega_d\left(\rho\right)\\ \Phi_{d,2,\boldsymbol{q}^{(2)}}\left(\rho\right)\propto\int_0^1 l_1^{q-1}\left(1-l_1\right)^{q-1}\Omega_d\left(\rho l_1\right)\Omega_d\left(\rho\left(1-l_1\right)\right)dl_1\\ \Phi_{d,n,\boldsymbol{q}^{(n)}}\left(\rho\right)\propto\int_0^1 l_1^{q-1}\left(1-l_1\right)^{(n-1)q-1}\Omega_d\left(\rho l_1\right)\Phi_{d,n-1,\boldsymbol{q}^{(n-1)}}\left(\rho\left(1-l_1\right)\right)dl_1\quad\left(n\geq 3\right)\end{cases}\tag{18}$$

We don’t have to worry about the proportionality constants in eq. 18, as their final values are simply obtained from the condition that $\Phi_{d,n,\boldsymbol{q}^{(n)}}\left(0\right)=1$ for any $n$. A walk of one step is, by definition, hyperspherical uniform, $HU\left(d,1,q,d\right)$, and its characteristic function $\Phi_{d,1}\left(\rho\right)$is $\Omega_d\left(\rho\right)$ for any $q>0$. Consistently, all hyperspace dimensions $k$ of table 1 reduce to $d$ for $n=1$but all Pearson-Dirichlet walks $PD\left(d,n,q\right)$are not hyperspherical uniform. We determine next the conditions for which the HU property holds for a walk of two steps $HU\left(d,2,q,k\right)$. Using eqs A-1and 18, it comes:

$$\Phi_{d,2,\boldsymbol{q}^{(2)}}\left(\rho\right)\propto\frac{1}{\rho^{2q-1}}\int_0^\rho x^{q-d/2}\left(\rho-x\right)^{q-d/2}J_{(d-2)/2}\left(x\right)J_{(d-2)/2}\left(\rho-x\right)dx\tag{19}$$

where $J_u\left(x\right)$ is a Bessel function of the first kind. The following finite integrals:

$$\int_0^\rho x^\mu\left(\rho-x\right)^\nu J_\mu\left(x\right)J_\nu\left(\rho-x\right)dx=\frac{\Gamma\left(\mu+1/2\right)\Gamma\left(\nu+1/2\right)}{\sqrt{2\pi}\Gamma\left(\nu+\mu+1\right)}\rho^{\mu+\nu+1/2}J_{\mu+\nu+1/2}\left(\rho\right)\tag{20}$$

$\left(\mu,\nu>-1/2\right)$ (integral 6.581.3 of [18]) and:

$$\int_0^{\rho} \frac{J_{\mu}(x) J_{\nu}(\rho-x)}{x(\rho-x)} dx = \frac{(\mu+\nu)}{\mu\nu\rho} J_{\mu+\nu}(\rho) \qquad (21)$$

$(\mu,\nu>0)$([19] p. 380) yield explicit expressions of integral eq. 19 either when $q=d-1$ (eq. 20, $\mu+\nu+1/2=(2d-3)/2$) or when $q=(d-2)/2$ (eq. 21, $\mu+\nu=d-2$). The parameters $q$ derived in section 3 (tables 1 and 3) are seen to obey the latter conditions. The walks of two steps, whose parameters are obtained from table 1 for $n=2$, are then concluded from eqs 19, 20, 21 and A-1 to be hyperspherical uniform with $k=2d-1$ for $q=d-1$ and $k=2d-2$ for $q=(d-2)/2$. For these walks, the HU property hold thus for $n=1,2$. Let us assume now that the walks with parameters $d,q,k$ given in table 1, are HU for $(n-1)\geq 3$ steps, that is $\Phi_{d,n-1,\boldsymbol{q}^{(n-1)}}(\rho)=\Omega_{a(n-1)+b}(\rho)$ where $a(=a(d))$ and $b$ are reported separately in table 3. Then, eq. 18 writes $(n\geq 2)$:

$$\Phi_{d,n,\boldsymbol{q}^{(n)}}(\rho) \propto \int_0^1 l_1^{q-1}(1-l_1)^{(n-1)q-1} \Omega_d(\rho l_1) \Omega_{a(n-1)+b}(\rho(1-l_1)) dl_1 \qquad (22)$$

which reduces to:

$$\Phi_{d,n,\boldsymbol{q}^{(n)}}(\rho) \propto \frac{1}{\rho^{nq-1}} \int_0^{\rho} x^{q-d/2} (\rho-x)^{(n-1)(q-a/2)-b/2} J_{(d\text{-}2)/2}(x) J_{(a(n\text{-}1)+b\text{-}2)/2}(\rho-x) dx \qquad (23)$$

From eq. 23 and integrals 20 and 21, it is deduced that, $\Phi_{d,n,\boldsymbol{q}^{(n)}}(\rho)=\Omega_{an+b}(\rho)$, for the walks whose parameters are given in tables 1 and 3. As the explicit calculations are all performed in the same way, we will just present one of them and derive the c.f., $\Phi_{2j+1,n,\boldsymbol{q}^{(n)}}(\rho)=\Omega_{(2j\text{-}1)n+2}(\rho)$, with $\left(\boldsymbol{q}^{(n)}=(2j-1)/2,..,(2j-1)/2\right)$ for a walk of $n$ steps in $\mathbb{R}^{2j+1}$. To obtain $\Phi_{2j+1,n,\boldsymbol{q}^{(n)}}(\rho)$, we assume that $\Phi_{2j+1,n-1,\boldsymbol{q}^{(n-1)}}(\rho)=\Omega_{(2j\text{-}1)(n-1)+2}(\rho)$ for a walk of $n-1$ steps $(n\geq 2)$ using the parameters of the third line of table 3, $a=2j-1$, $b=2$. From the recurrence relation (eq. 23), it comes:

$$\Phi_{2j+1,n,\boldsymbol{q}^{(n)}}(\rho) \propto \frac{1}{\rho^{(2j-1)n/2-1}} \int_0^{\rho} x^{-1} (\rho - x)^{-1} J_{(2j-1)/2}(x) J_{(2j-1)(n-1)/2}(\rho - x) dx \qquad (24)$$

From eq. 21 with $\mu = (2j-1)/2$ and $\nu = (2j-1)(n-1)/2$), we deduce that $\Phi_{2j+1,n,\boldsymbol{q}^{(n)}}(\rho) \propto J_{(2j-1)n/2}(x) / \rho^{(2j-1)n/2}$ and from eq. A-1 and the condition, $\Phi_{2j+1,n,\boldsymbol{q}^{(n)}}(0) = 1$, we obtain finally $\Phi_{2j+1,n,\boldsymbol{q}^{(n)}}(\rho) = \Omega_{(2j-1)n+2}(\rho)$ which proves that the HU property holds for $n$ when it holds for $n-1$ $(n \geq 2)$.

In sum, we have shown that any walk of $n$ steps defined in table 1 is HU given that it is HU for $(n-1)$steps $(n \geq 2)$. As the property holds for $n = 1$(and for $n = 2$) it holds for any $n$. We conclude that the walks evidenced by the necessary condition of section 3 are all hyperspherical for any $n$. The corresponding parameters and distributions of the endpoint are given in table 1. Two families of HU Dirichlet walks are seen to exist in any space $\mathbb{R}^d$ with $d \geq 3$ and only one family for $d = 2$.

When $q$ is an integer , a walk $HU(d,n,q,k)$ can be interpreted equally in term of a random walk similar to the walk described in the introduction [1] (figure 1): instead of changing its direction after every step with an exponentially distributed length, the particle changes it at every $q$ steps, the intermediate steps being ineffective (figure 1). The distribution of the step length $s$ between two changes of direction is then the sum of $q$ i.i.d. exponential random contributions, that is a gamma distribution, $p(s) = s^{q-1} \mathrm{e}^{-s} / \Gamma(q)$. Equivalently, we may consider that the $n$ steps of the walk are i.i.d. $d$-dimensional random vectors whose lengths have a gamma distribution with a pdf $p(s) = s^{q-1} \mathrm{e}^{-s} / \Gamma(q)$, where $q$ may have any positive value, and whose orientations are uniform. The latter walk, given its total length being equal to 1, is a Pearson-Dirichlet walk whose step lengths have by definition a Dirichlet distribution $D_m(\boldsymbol{q}^{(n)})$ (section 2). The distribution of the distance from the origin to the endpoint of an unconditioned walk of $n$ i.i.d. steps in $\mathbb{R}^d$ $(d \geq 2)$ with uniform orientation and a step length gamma distribution, $p(s) = s^{d-2} \mathrm{e}^{-s} / (d-2)!$ , is given in section 8 (eq. 47).

Similarly, the conditional pdf of the times of occurrence of Poisson events in $(0,1]$, given their number $N(1)=nq-1$, is $p\left(t_1,t_2,..,t_{nq-1}\left|N(1)=nq-1\right.\right)=(nq-1)!$ $\left(0<t_1<t_2<...<t_{nq-1}\le 1\right)$ (section 2, [7-8]) and the inter-arrival times distribution is thus a Dirichlet distribution $D_{nq-1}\left(\boldsymbol{\alpha}^{(nq)}=(1,1,...,1)\right)$. Amalgamating the $nq$ variables $q$ by $q$ (figure 1, section 2) gives the sought-after Dirichlet distribution $D_m\left(\boldsymbol{q}^{(n)}\right)$. When applied to a random flight similar to that investigated in [4-5], the previous discussion means that the particle changes its direction at every $q$ Poisson events, $q-1$ intermediate events being ineffective (figure 1).

## 5. Additional hyperspherical uniform walks

Two other finite integrals of products of powers and Bessel functions of the first kind yield additional HU walks.

A HU walk of two steps in any space of dimension $d$ greater than 1 is indeed obtained for the following Dirichlet distribution $D_1\left(\boldsymbol{\alpha}^{(2)}=(d-1,d)\right)$(and by symmetry $D_1\left(\boldsymbol{\alpha}^{(2)}=(d,d-1)\right)$):

$$p_1(l_1)=\frac{(2d-2)!}{(d-1)!(d-2)!}l_1^{d-2}(1-l_1)^{d-1} \tag{25}$$

Then, from $\boldsymbol{r}_2^{(d)}=l_1\boldsymbol{u}_1^{(d)}+(1-l_1)\boldsymbol{u}_2^{(d)}$, we write the characteristic function:

$$\Phi_{d,2,(d-1,d)}(\rho)=\left\langle\exp\left(i\boldsymbol{\rho}.\boldsymbol{r}_2^{(d)}\right)\right\rangle\propto\int_0^1 l_1^{d-2}(1-l_1)^{d-1}\,\Omega_d(\rho l_1)\Omega_d\left(\rho(1-l_1)\right)dl_1 \tag{26}$$

that is :

$$\Phi_{d,2,(d-1,d)}(\rho)\propto\frac{1}{\rho^{2d-2}}\int_0^{\rho}x^{d/2-1}(\rho-x)^{d/2}\,J_{d/2-1}(x)J_{d/2-1}(\rho-x)dx \tag{27}$$

From $\Phi_{d,2,(d-1,d)}(0)=1$ and the following integral:

$$\int_0^{\rho} x^{\mu}(\rho-x)^{\nu+1} J_{\mu}(x) J_{\nu}(\rho-x)\,dx=\frac{\Gamma(\mu+1/2)\Gamma(\nu+3/2)}{\sqrt{2\pi}\Gamma(\nu+\mu+2)}\rho^{\mu+\nu+3/2} J_{\mu+\nu+1/2}(\rho) \qquad (28)$$

$(\mu>-1/2, \nu>-1)$ (integral 6.581.4 of [18]), and from $\mu+\nu+1/2=(2d-3)/2$, we get (eq. A-1):

$$\Phi_{d,2,(d-1,d)}(\rho)=\Omega_{2d\text{-}1}(\rho) \qquad (29)$$

The latter walk is then concluded to be HU with $k=2d-1$. The distribution of the endpoint is finally:

$$p_{d,2,\{d-1,d\}}(\boldsymbol{r})=\frac{\Gamma((2d-1)/2)}{\Gamma((d-1)/2)\pi^{d/2}}\left(1-r^2\right)^{(d-3)/2} \qquad (30)$$

and consequently:

$$P_{d,2,\{d-1,d\}}(r)=\frac{2^{d-1}\Gamma((2d-1)/2)}{(d-2)!\sqrt{\pi}} r^{d-1}\left(1-r^2\right)^{(d-3)/2} \qquad (31)$$

A third HU walk, whose endpoint is uniformly distributed in the inside of a sphere in $\mathbb{R}^3$, is then found for a walk of two steps.

The results of the previous paragraph extend to a walk in $\mathbb{R}^d$ whose step lengths have a Dirichlet distribution $D_m\left(\boldsymbol{\alpha}^{(n)}=(d,d-1,d-1,..,d-1)\right)$. Except for the first, the Dirichlet parameters of that walk coincide with those of the walks defined by the second line of table 1. The walk $D_m\left(\boldsymbol{\alpha}^{(n)}\right)$ is hyperspherical uniform with a hyperspace dimension, $k=n(d-1)+1$, that coincides too with those of the aforementioned walk.

A last family of HU walks is found in any space of dimension $d$ greater than 2 from the following integral $(\mu>0, \nu>-1)$( [19] p. 380):

$$\int_0^{\rho} \frac{J_\mu(x) J_v(\rho - x)}{x} dx = \frac{1}{\mu} J_{\mu+v}(\rho) \tag{32}$$

The associated Dirichlet distribution is, $D_1\left(\boldsymbol{\alpha}^{(2)} = \left(\frac{d}{2} - 1, \frac{d}{2}\right)\right)$(and by symmetry $D_1\left(\boldsymbol{\alpha}^{(2)} = \left(\frac{d}{2}, \frac{d}{2} - 1\right)\right)$), for a two-step walk with a distribution of $l_1$ given by:

$$p_1(l_1) = \frac{(d-2)!}{\Gamma((d-2)/2)\Gamma(d/2)} l_1^{(d-4)/2} (1 - l_1)^{(d-2)/2} \tag{33}$$

As above, we write the characteristic function from $\boldsymbol{r}_2^{(d)} = l_1 \boldsymbol{u}_1^{(d)} + (1\text{-}l_1)\boldsymbol{u}_2^{(d)}$:

$$\Phi_{d,2,(d/2-1,d/2)}(\rho) \propto \frac{1}{\rho^{d\text{-}2}} \int_0^{\rho} \frac{J_{(d\text{-}2)/2}(x) J_{(d\text{-}2)/2}(\rho - x)}{x} dx \tag{34}$$

and we get from eqs 32 and A-1:

$$\Phi_{d,2,(d/2-1,d/2)}(\rho) = \Omega_{2d\text{-}2}(\rho) \tag{35}$$

from which the latter walk is concluded to be HU with $k = 2d - 2$. Similarly, we obtain that the c.f. of a walk of three steps, with a step length distribution given by $D_2\left(\boldsymbol{\alpha}^{(3)} = \left(\frac{d}{2}, \frac{d}{2} - 1, \frac{d}{2} - 1\right)\right)$, writes:

$$\Phi_{d,3,\{d/2,d/2-1,d/2-1\}}(\rho) \propto \frac{1}{\rho^{3(d-2)/2}} \int_0^{\rho} \frac{J_{(d\text{-}2)/2}(x) J_{d\text{-}2}(\rho - x)}{x} dx \propto \frac{J_{3(d\text{-}2)/2}(\rho)}{\rho^{3(d-2)/2}} \tag{36}$$

That is, $\Phi_{d,3,\{d/2,d/2-1,d/2-1\}}(\rho) = \Omega_{3d-4}(\rho)$. As before, these results hold for any walk in $\mathbb{R}^d$ $(d > 2)$ whose step lengths have a Dirichlet distribution

$D_m\left(\boldsymbol{\alpha}^{(n)}=\left(d/2, d/2-1, d/2-1,.., d/2-1\right)\right)$ and $k=n(d-2)+2$. The latter walk is then HU with a hyperspace dimension which coincides with that of the walk of the third line of table 1. The distribution of the endpoint and that of the distance from the endpoint to the origin are therefore given by table 1 with $\delta=\left(n(d-2)-d\right)/2$. Thus, two uniform walks, given by $n=d/(d-2)$, are obtained for the couples $(d=3, n=3)$ with $\boldsymbol{\alpha}^{(3)}=\left(\frac{3}{2},\frac{1}{2},\frac{1}{2}\right)$ and $(d=4, n=2)$ with $\boldsymbol{\alpha}^{(2)}=(2,1)$. As seen in table 2, these walks occur in the same spaces with the same number of steps than two previous walks but their step length distributions are different.

In sum, five additional 'uniform' walks were found in the present work: two walks of two steps and two of three steps in $\mathbb{R}^3$ and one of two steps in $\mathbb{R}^4$. Table 2 collects the characteristics of the seven uniform walks which belong to the investigated Pearson-Dirichlet family. The uniform walks found in $\mathbb{R}^3$ and in $\mathbb{R}^4$ are seen to be degenerate as they are obtained for more than one set of Dirichlet parameters.

## 6. Stochastic representation of HU walks

The HU property results in a simple stochastic representation of the endpoint $\boldsymbol{r}_n^{(d)}$ of a $HU(d,n,q,k)$ walk. A $k$-dimensional Gaussian vector $\boldsymbol{G}^{(k)}$, $N(\boldsymbol{0},\boldsymbol{I}_k)$ with $\boldsymbol{I}_k$ the unit $k\times k$ matrix, whose components are independent random variables with zero means and variances of 1, yields, when normalized, a unit vector $\boldsymbol{u}^{(k)}=\boldsymbol{G}^{(k)}/\left\|\boldsymbol{G}^{(k)}\right\|$ whose tip is uniformly distributed over the surface of the unit hypersphere in $\mathbb{R}^k$ (appendix). The square of the modulus of $\boldsymbol{G}^{(k)}$, $\chi_k^2=\left\|\boldsymbol{G}^{(k)}\right\|^2$, follows, by definition, a chi-square distribution with $k$ degrees of freedom [13]. It is too a gamma distribution with a shape parameter of $k/2$ and a scale parameter of $1/2$. It can be split into two independent chi-squared random variables: $\chi_k^2=\chi_d^2+\chi_{k-d}^2$ with respective degrees of freedom $d$ and $k-d$. To obtain the endpoint of the walk $\boldsymbol{r}_n^{(d)}$, it suffices to take the first $d$ components of $\boldsymbol{u}^{(k)}$. Therefore:

$$\boldsymbol{r}_n^{(d)} \triangleq \frac{\boldsymbol{G}^{(d)}}{\sqrt{\left\|\boldsymbol{G}^{(d)}\right\|^2 + \chi_{k-d}^2}} \tag{37}$$

where $\boldsymbol{G}^{(d)}$ is now a $d$ -dimensional Gaussian vector, $N\left(\boldsymbol{0}, \boldsymbol{I}_d\right)$, where $a \triangleq b$ means that $a$ and $b$ have the same distribution and where, by convention, $\chi_0^2 = 0$. The vector $\boldsymbol{G}^{(d)}$ and the random variable $\chi_{k-d}^2$ are independent. Similarly, the distance of the endpoint to the origin is represented by:

$$r_n^{(d)} \triangleq \frac{\chi_d}{\sqrt{\chi_d^2 + \chi_{k-d}^2}} \tag{38}$$

( $\chi_k$ has a chi distribution [13] with $k$ degrees of freedom). Equations 37 and 38 may be used to perform fast Monte-Carlo simulations of any $HU\left(d,n,q,k\right)$ walk. It suffices indeed to generate $d+1$ independent random variables to simulate the endpoint positions $\boldsymbol{r}_n^{(d)}$ and only two to simulate $r_n^{(d)}$ for any $n$ and $d$ .

**7. Asymptotic behavior**

For $n=1$, the endpoints of any walk $PD\left(d,n,q\right)$ are uniformly distributed over the surface of a unit $d$ -dimensional hypersphere. When $n$ increases for a fixed $d$ , the endpoints invade progressively the inner part of the hypersphere forming a spherically symmetric cloud for any $n$. When $n \to \infty$, the latter cloud shrinks gradually into a Gaussian spherical cloud which is more and more concentrated around the origin. In all cases, $\left\langle \left\|\boldsymbol{r}_n^{(d)}\right\|^2 \right\rangle = d\left\langle r_1^2 \right\rangle = \frac{q+1}{\left(nq+1\right)}$ (eq. 12), decreases regularly with $n$ independently of $d$ . For the $HU\left(d,n,q,k\right)$ walks, the latter scenario is a direct consequence of a theorem of Diaconis and Freedman [20] which proves that the first $d$ coordinates of a point uniformly distributed over the surface of a $k = an+b$ sphere are independent standard normal variables, in the limit as $n \to \infty$ with $d$ fixed.

For any dimension $d$ , the main term which contributes to the moment $\left\langle r_1^{2p} \right\rangle$ of a single component of $\boldsymbol{r}_n^{(d)}$ in the limit as $n \to \infty$ , is $\frac{(2p)!}{2^p} \times \frac{n(n-1)..(n-p+1)}{p!} \times \left\langle l_1^2 l_2^2 .. l_p^2 \right\rangle \times \left\langle u_1^2(1) \right\rangle^p$.

From eqs 8 and 10, it comes:

$$\left\langle r_1^{2p} \right\rangle_\infty = \lim_{n \to \infty} \frac{(2p)!}{2^p d^p} \times \frac{n(n-1)..(n-p+1)}{p!} \times \frac{\left(q(q+1)\right)^p}{\prod_{j=1}^{2p}(nq+j-1)} = (2p-1)!! \left( \frac{q+1}{nqd} \right)^p \qquad (39)$$

which are the moments of a Gaussian distribution with a zero mean and a variance equal to $\frac{q+1}{nqd}$. As the distribution of $\boldsymbol{r}_n^{(d)}$ is spherically symmetric, the latter argument indicates that the Gaussian behavior holds in the asymptotic limit for any walk $PD(d, n \to \infty, q)$ with $d$ fixed.

## 8. Pearson-Liouville random walks

### 8.1 Definition and generalities

The Pearson-Dirichlet walk, and more particularly the walks $HU(d,n,q,k)$ whose parameters are given in table 1, can serve as "unit" walks to mix walks with different total lengths $l$. We assume then that the total length $l$ is distributed according to some continuous probability density function $f(l)$ and we denote the new step lengths as $\boldsymbol{s}^{(n)} = (s_1,..,s_n)$ $\left( \sum_{k=1}^{n} s_k = l \right)$. The renormalized step lengths, $\boldsymbol{v}^{(n)} = \left(l_1 = s_1/l,..,l_n = s_n/l\right)$ $\left( \sum_{k=1}^{n} l_k = 1 \right)$ have a Dirichlet distribution $D_m\left(\boldsymbol{q}^{(n)}\right)$ for any value of $l$. Then the joint pdf of $\boldsymbol{s}^{(n)}$ is:

$$p_L(s_1,..,s_n) = \frac{\Gamma(nq)}{\Gamma(q)^n} \times \prod_{k=1}^{n} s_k^{q-1} \times \frac{f\left(\sum_{k=1}^{n} s_k\right)}{\left(\sum_{k=1}^{n} s_k\right)^{nq-1}} \qquad (40)$$

The step length distribution, given by eq. 40, is a Liouville distribution with a generating density $f(.)$ (chapter 6 of [12], [21]). The associated random walk will be named consistently a Pearson-Liouville random walk. Its stochastic representation is $\boldsymbol{s}^{(n)} \triangleq l\boldsymbol{v}^{(n)}$, where $l$ and $\boldsymbol{v}^{(n)}$ are independent. If the generating density is defined in a finite interval $(0,L)$, then $p_L(s_1,..,s_n)$ is defined in the simplex $\left((s_1,..,s_n):\sum_{k=1}^{n} s_k \le L\right)$. If $f(.)$ is defined on $\mathbb{R}^+$, the marginal distribution of a step length $s_k$ $(k=1,..,n)$ is calculated from that of a Dirichlet random walk (section 2) to be:

$$p(s_k) = \frac{\Gamma(nq)}{\Gamma(q)\Gamma((n-1)q)} \int_{s_k}^{\infty} f(l) \left(\frac{s_k}{l}\right)^{q-1} \left(1-\frac{s_k}{l}\right)^{(n-1)q-1} \frac{dl}{l} \tag{41}$$

The probability density function of the endpoint, $g_{d,n,q}(\boldsymbol{r})$, and that, $G_{d,n,q}(r)$, of the distance from the origin to the endpoint of a Pearson-Liouville walk associated with the HU Dirichlet walks read then:

$$\begin{cases} I_{d,n,q}(r) = \dfrac{1}{\Gamma(\delta+1)} \displaystyle\int_{r}^{\infty} \frac{f(l)}{l^{d+2\delta}} \left(l^2 - r^2\right)^{\delta} dl \\ g_{d,n,q}(\boldsymbol{r}) = \left(\Gamma(k/2)/\pi^{d/2}\right) I_{d,n,q}(r) \\ G_{d,n,q}(r) = 2\left(\Gamma(\delta+1+d/2)/\Gamma(d/2)\right) r^{d-1} I_{d,n,q}(r) \end{cases} \tag{42}$$

as deduced from eq. 16 with $k$ and $\delta$ given in table 1. The endpoint $\boldsymbol{r}_n^{(d)}$ of such a Pearson-Liouville walk is the projection in $\mathbb{R}^d$ of a vector $\boldsymbol{r}_n^{(k)}$ of $\mathbb{R}^k$. Its stochastic representation is $\boldsymbol{r}_n^{(k)} \triangleq l\boldsymbol{u}^{(k)}$, where $l$ is independent of $\boldsymbol{u}^{(k)}$ which is a unit vector whose tip is uniformly distributed over the surface of the unit hypersphere in $\mathbb{R}^k$. Therefore, we get from eq. 37 that:

$$\boldsymbol{r}_n^{(d)} \triangleq \frac{l\boldsymbol{G}^{(d)}}{\sqrt{\left\|\boldsymbol{G}^{(d)}\right\|^2 + \chi_{k-d}^2}} \tag{43}$$

where $l$ is of course independent of the random vector $\boldsymbol{G}^{(d)}$ and of the chi-square $\chi^2_{k-d}$. A stochastic equation for the distance from the origin to the endpoint might similarly be written from eq. 38. The previous representations are efficient for Monte-Carlo simulations of the endpoints of these walks. If we define now a vector $\boldsymbol{t}^{(k)}$ of $\mathbb{R}^k$, whose distribution is spherical with a density $p_k\left(t=\left\|\boldsymbol{t}^{(k)}\right\|\right)$ equal to the pdf $f(t)$ normalized by the area of the hypersphere of radius $t$ in $\mathbb{R}^k$:

$$p_k(t)=\frac{\Gamma(k/2)\,f(t)}{2\pi^{k/2}t^{k-1}} \tag{44}$$

Then the $d$-dimensional projection $\boldsymbol{r}=\boldsymbol{r}_n^{(d)}$ of $\boldsymbol{t}^{(k)}$ has a density $g_{d,n,q}(\boldsymbol{r})$. The latter result is consistently found by a direct application of relation 28 of [22]. The characteristic function of the distribution of $\boldsymbol{r}$ depends only on the modulus $\rho=\|\boldsymbol{\rho}\|$ as $g_{d,n,q}(\boldsymbol{r})$ is spherically symmetric. It reads:

$$\Phi_{d,n,q}(\rho)=\left\langle e^{i\boldsymbol{\rho}.\boldsymbol{r}}\right\rangle=\frac{2^{(k-2)/2}\Gamma(k/2)}{\rho^{(k-2)/2}}\int_0^\infty\frac{f(l)\,J_{(k-2)/2}(\rho l)}{l^{(k-2)/2}}dl \tag{45}$$

The pdf's $g_{d,n,q}(\boldsymbol{r})$ and $G_{d,n,q}(r)$ (eq. 42) can be derived alternatively from the characteristic function by the following inversion formula (eqs 6 and 10 of [22]):

$$\left\{\begin{aligned} M_{d,n,q}(r)&=\int_0^\infty\rho^{d/2}J_{(d-2)/2}(r\rho)\Phi_{d,n,q}(\rho)\;d\rho\\ g_{d,n,q}(\boldsymbol{r})&=M_{d,n,q}(r)\Big/\left([2\pi]^{d/2}\,r^{(d-2)/2}\right)\\ G_{d,n,q}(r)&=r^{d/2}M_{d,n,q}(r)\Big/\left(2^{(d-2)/2}\Gamma(d/2)\right)\end{aligned}\right. \tag{46}$$

Eq. 42 is consistently obtained when plugging eq. 45 into eq. 46, interchanging the order of integration and using integral 6.575.1 of [18].

8.2 An example: i.i.d. gamma distributed step lengths

When the distribution of $l$ is chosen to be a gamma distribution, with a pdf $f(l) = l^{nq-1}\exp(-l)/\Gamma(nq)$, then the pdf $p_L(s_1,..,s_n)$ (eq. 40) is, as expected, that of $n$ i.i.d. gamma random variables, with $p(s_k) = s_k^{q-1}\exp(-s_k)/\Gamma(q)$ $(k = 1,..,n)$, a result which is readily obtained from eq. 41.

For the first HU family, the parameters are $q = d-1$ and $\delta = (n(d-1)-(d+1))/2$ (second line of table 1). The pdf of the endpoint $g_{d,n,d-1}(\boldsymbol{r})$ and the pdf of the distance $G_{d,n,d-1}(r)$ of a walk of $n \geq 2$ i.i.d. steps in $\mathbb{R}^d$ $(d \geq 2)$, are then calculated from eq. 42 using integral 3.387.6 of [18]. These pdf's and the characteristic function of the distribution $g_{d,n,d-1}(\boldsymbol{r})$ calculated from eq. 45 are:

$$\begin{cases} \nu = (n(d-1)-d)/2 \\ F_{d,n}(r) = r^{\nu}K_{\nu}(r)/\left(2^{\nu+d-1}\Gamma(n(d-1)/2)\right) \\ g_{d,n,d-1}(\boldsymbol{r}) = F_{d,n}(r)/\pi^{d/2} \\ G_{d,n,d-1}(r) = 2r^{d-1}F_{d,n}(r)/\Gamma(d/2) \\ \Phi_{d,n,d-1}(\rho) = 1/\left(1+\rho^2\right)^{n(d-1)/2} \end{cases} \quad (47)$$

where $K_{\nu}(x)$ is a modified Bessel function of the second kind. The density, $g_{d,n,d-1}(r)$ and $\Phi_{d,n,d-1}(\rho)$, once properly normalized, are dual spherical densities [23-24] related through Hankel transforms [22]. The latter is that of a *d*-dimensional spherical Student distribution with $\nu$ degrees of freedom [12].

For the second HU family, with $q = (d-2)/2$ $(d \geq 3)$, $n \geq 2$ and $\delta = (n(d-2)-d)/2$ (third line of table 1), the endpoint, the distance pdf's and the characteristic function (eq. 45) read:

$$\begin{cases} I_{n,d}(r) = \dfrac{n(d-2)}{\Gamma((n-1)(d-2)/2)} \displaystyle\int_r^{\infty} \dfrac{\exp(-l)}{l^{n(d-2)/2+1}} \left(l^2 - r^2\right)^{(n(d-2)-d)/2} dl \\ g_{d,n,(d-2)/2}(\mathbf{r}) = I_{n,d}(r) \big/ \left(2\pi^{d/2}\right) \\ G_{d,n,(d-2)/2}(r) = r^{d-1} I_{n,d}(r) / \Gamma(d/2) \\ \Phi_{d,n,(d-2)/2}(\rho) = \left[2\left(\sqrt{1+\rho^2} - 1\right) \big/ \rho^2\right]^{n(d-2)/2} \end{cases} \qquad (48)$$

Explicit general solutions, like those given by eq. 47, were not found in that case. Precise numerical calculations of $G_{d,n,(d-2)/2}(r)$ can however be performed from eqs 46 and 48. The latter distribution can be obtained in some specific cases, for instance for $d = 3$ and $n = 3,5$:

$$\begin{cases} G_{3,3,1/2}(r) = 8r^2 erfc\left(\sqrt{r}\right) + \dfrac{2}{\sqrt{\pi}} \sqrt{r}\,(2-4r)\exp(-r) \\ G_{3,5,1/2}(r) = \dfrac{4}{3} r^2 \left(4r^2 - 15\right) erfc\left(\sqrt{r}\right) + \dfrac{8}{3\sqrt{\pi}} r^{3/2} \left(6 + r - 2r^2\right)\exp(-r) \end{cases} \qquad (49)$$

where $erfc(x)$ is the complementary error function.

Many of the Pearson-Dirichlet walks $HU(d,n,q,k)$, and in particular all uniform walks, and the Pearson-Liouville walks defined in section 8.2 were further investigated by Monte-Carlo simulations to obtain “experimental” distributions of the distance of the endpoint to the origin. All results were found to be in excellent agreement with the corresponding closed-form distributions derived in the present work. It is worth mentioning that important aspects of the considered walks are connected with the general problem of the random fragmentation of the unit interval (see for instance [25] and references therein). Finally, there is a deep connection between HU walks and finite integrals of products of Bessel functions and powers.

**Acknowledgments:**

I thank Renaud Delannay (Université de Rennes I) for useful discussions.

## Appendix:

## Uniform distribution of a vector over the surface of the unit hypersphere in $\mathbb{R}^N$

Consider first a unit vector $\boldsymbol{u}^{(N)}$ whose tip spans uniformly the surface of the hypersphere in $\mathbb{R}^N$. Using hyperspherical coordinates, the characteristic function (c.f.) $\Omega_N(\boldsymbol{\rho}) = \left\langle e^{i\boldsymbol{\rho}.\boldsymbol{u}^{(N)}} \right\rangle$ of the distribution of $\boldsymbol{u}^{(N)}$, which depends only on the modulus $\rho = \|\boldsymbol{\rho}\|$, is found to be:

$$\Omega_N(\rho) = \frac{\Gamma(N/2)}{\sqrt{\pi}\,\Gamma((N-1)/2)} \int_0^{\pi} e^{i\rho\cos(\theta)} \sin^{N-2}(\theta)\, d\theta = \frac{2^{(N-2)/2}\,\Gamma(N/2)}{\rho^{(N-2)/2}} J_{(N-2)/2}(\rho) \qquad \text{(A-1)}$$

where the c.f. has been expressed in term of a Bessel function of the first kind, $J_u(\rho)$. More generally, the c.f. $\Phi_N(\boldsymbol{\rho}) = \left\langle e^{i\boldsymbol{\rho}.\boldsymbol{r}^{(N)}} \right\rangle$ of any spherically symmetric vector $\boldsymbol{r}^{(N)}$ of $\mathbb{R}^N$, whose distribution is invariant by any orthogonal transformation, is similarly a function of the sole modulus of $\boldsymbol{\rho}$ [12]. Thus, the c.f. of the marginal distribution of any number $j\,(j = 1,..,N)$ of components of a spherical vector $\boldsymbol{r}^{(N)}$ is still $\Phi_N(\rho)$ but $\rho$ is now the modulus of a vector $\boldsymbol{\rho}$ in which $N - j$ components are made equal to zero. To prove that a spherically symmetric vector $\boldsymbol{r}_n^{(d)}$ is the projection of a unit vector $\boldsymbol{u}^{(N)}$, it suffices thus to show that the c.f. of the first component $r_1$ of $\boldsymbol{r}_n^{(d)}$ is $\Phi_1(\rho) = \left\langle e^{i\rho.r_1} \right\rangle = \Omega_N(\rho)$.

Consider now a vector $\boldsymbol{G}^{(N)} = (G_1,..,G_N)$ of $\mathbb{R}^N$, whose components are i.i.d. standard Gaussian variables with a zero mean and a variance of 1 and whose modulus is $G = \left\|\boldsymbol{G}^{(N)}\right\|$. Then the tip of the unit vector $\boldsymbol{u}^{(N)} = (G_1/G,..,G_N/G)$ spans uniformly the surface of the hypersphere in $\mathbb{R}^N$ (see for instance [12] page 20). Every $G_i^2$ $(i = 1,..,N)$ is gamma distributed with a shape parameter of $1/2$ and a scale parameter of $1/2$ (it has a chi distribution with one degree of freedom [13]). The distribution of $\left(u_1^2 = G_1^2/G^2,..,u_N^2 = G_N^2/G^2\right)$ is consequently a Dirichlet distribution whose parameters are all equal to 1/2. Then, the joint distribution of any number $j$ components of $\boldsymbol{u}^{(N)}$ can be obtained by using the amalgamation property. Its pdf is given by [12, 17]:

$$p_N\left(u_1,u_2,..,u_j\right)=\frac{\Gamma\left(N/2\right)}{\Gamma\left(\left(N-j\right)/2\right)\pi^{j/2}}\left(1-\sum_{i=1}^{j}u_i^2\right)^{\left(N-j-2\right)/2}\quad\left(\sum_{i=1}^{j}u_i^2<1\right)\qquad\left(\text{A-2}\right)$$

The previous Dirichlet distribution, $D_{N-1}\left(\boldsymbol{\alpha}^{(N)}=\left(1/2,..,1/2,1/2\right)\right)$, of $\left(u_1^2,..,u_N^2\right)$ and eq. 8 with $\boldsymbol{\alpha}^{(N)}$ and $\boldsymbol{\beta}^{(N)}=\left(p,0,...,0\right)$ yield the even moments $\left\langle u_i^{2p}\right\rangle$ of a single component of $\boldsymbol{u}^{(N)}$:

$$\left\langle u_i^{2p}\right\rangle=\frac{\left(1/2\right)_p}{\left(N/2\right)_p}=\frac{\left(2p-1\right)!!}{\prod_{j=1}^{p}\left(N+2j-2\right)}\qquad\left(\text{A-3}\right)$$

Table 1 :

Parameters needed to obtain the pdf of the endpoint $p_{d,n,q}(\boldsymbol{r})$, and that of the distance of the endpoint to the origin $P_{d,n,q}(r)$, for all Pearson-Dirichlet hyperspherical uniform walks of $n(\geq 2)$ steps in $\mathbb{R}^d$ whose step length distributions are $D_m\left(\boldsymbol{q}^{(n)} = (q, q, .., q)\right)$

| Parameter of the Dirichlet distribution: (condition: $q(d) > 0$) $q$ | Hypersphere in $\mathbb{R}^k$: $k$ | $p_{d,n,q}(\boldsymbol{r}) \propto \left(1-r^2\right)^{\delta}$ exponent $\delta = (k-d-2)/2$ |
|---|---|---|
| $d-1$ | $n(d-1)+1$ | $\frac{n(d-1)-(d+1)}{2}$ |
| $\frac{d}{2}-1$ | $n(d-2)+2$ | $\frac{n(d-2)-d}{2}$ |
| $p_{d,n,q}(\boldsymbol{r}) = \frac{\Gamma(k/2)}{\Gamma(\delta+1)\pi^{d/2}}\left(1-r^2\right)^{\delta}$ | $P_{d,n,q}(r) = \frac{2\Gamma(\delta+d/2+1)}{\Gamma(\delta+1)\Gamma(d/2)} r^{d-1}\left(1-r^2\right)^{\delta}$ | |

Table 2 : The seven Pearson-Dirichlet hyperspherical uniform walks, with step length distributions $D_m\left(\boldsymbol{\alpha}^{(n)}=(\alpha_1,\alpha_2,..,\alpha_n)\right)$, whose endpoints are uniformly distributed in the inside of a hypersphere in $\mathbb{R}^d$ ( the first and the sixth walks are described in [1,4-5]).

| Walk in $\mathbb{R}^d$ | Number of steps $n$ | Dirichlet parameters $\boldsymbol{\alpha}^{(n)}=(\alpha_1,\alpha_2,..,\alpha_n)$ | Step length pdf's $D_{n\text{-}1}\left(\boldsymbol{\alpha}^{(n)}\right)$ |
|---|---|---|---|
| $\mathbb{R}^2$ | 3 | $(1,1,1)$ | $p_2(l_1,l_2)=2 \quad (l_3=1\text{-}l_1-l_2)$<br>$p_1(l_i)=2(1-l_i) \quad (i=1,.,3)$ |
| $\mathbb{R}^3$ | 2 | $(2,2)$ | $p_1(l_1)=6l_1(1-l_1) \quad (l_2=1\text{-}l_1)$ |
| $\mathbb{R}^3$ | 2 | $(2,3)$ | $p_1(l_1)=12l_1(1-l_1)^2 \quad (l_2=1\text{-}l_1)$ |
| $\mathbb{R}^3$ | 3 | $\left(\frac{1}{2},\frac{1}{2},\frac{1}{2}\right)$ | $p_2(l_1,l_2)=1\big/\left(2\pi\sqrt{l_1l_2l_3}\right)$<br>$(l_3=1\text{-}l_1-l_2)$<br>$p_1(l_i)=1\big/\left(2\sqrt{l_i}\right) \quad (i=1,.,3)$ |
| $\mathbb{R}^3$ | 3 | $\left(\frac{3}{2},\frac{1}{2},\frac{1}{2}\right)$ | $p_2(l_1,l_2)=(3/(2\pi))\times\sqrt{l_1/(l_2l_3)}$<br>$(l_3=1\text{-}l_1-l_2)$<br>$p_1(l_1)=3\sqrt{l_1}/2$<br>$p_1(l_i)=3(1-l_i)\big/\left(4\sqrt{l_i}\right) \quad (i=2,3)$ |
| $\mathbb{R}^4$ | 2 | $(1,1)$ | $p_1(l_1)=1 \quad (l_2=1\text{-}l_1)$ |
| $\mathbb{R}^4$ | 2 | $(2,1)$ | $p_1(l_1)=2l_1 \quad (l_2=1\text{-}l_1)$ |

Table 3 : The parameters of the Pearson-Dirichlet hyperspherical uniform walks in $\mathbb{R}^d$ , with step length distributions $D_m\left(\boldsymbol{q}^{(n)}=(q,q,..,q)\right)$, which are needed to solve the recurrence relations (eqs 22-23) from integrals eqs 20 and 21

(hypersphere in $\mathbb{R}^k$ with $k=a(n-1)+b$ for $n-1$ steps)

| $q\ (>0)$ | $a$ | $b$ | Integral eq. no | $\mu$ | $\nu$ $(n-1 \text{ steps})$ |
|---|---|---|---|---|---|
| $d-1$ | $d-1$ | 1 | 20 | $\frac{d}{2}-1$ | $\frac{(d-1)(n-1)-1}{2}$ |
| $\frac{d}{2}-1$ | $d-2$ | 2 | 21 | $\frac{d}{2}-1$ | $\frac{(d-2)(n-1)}{2}$ |

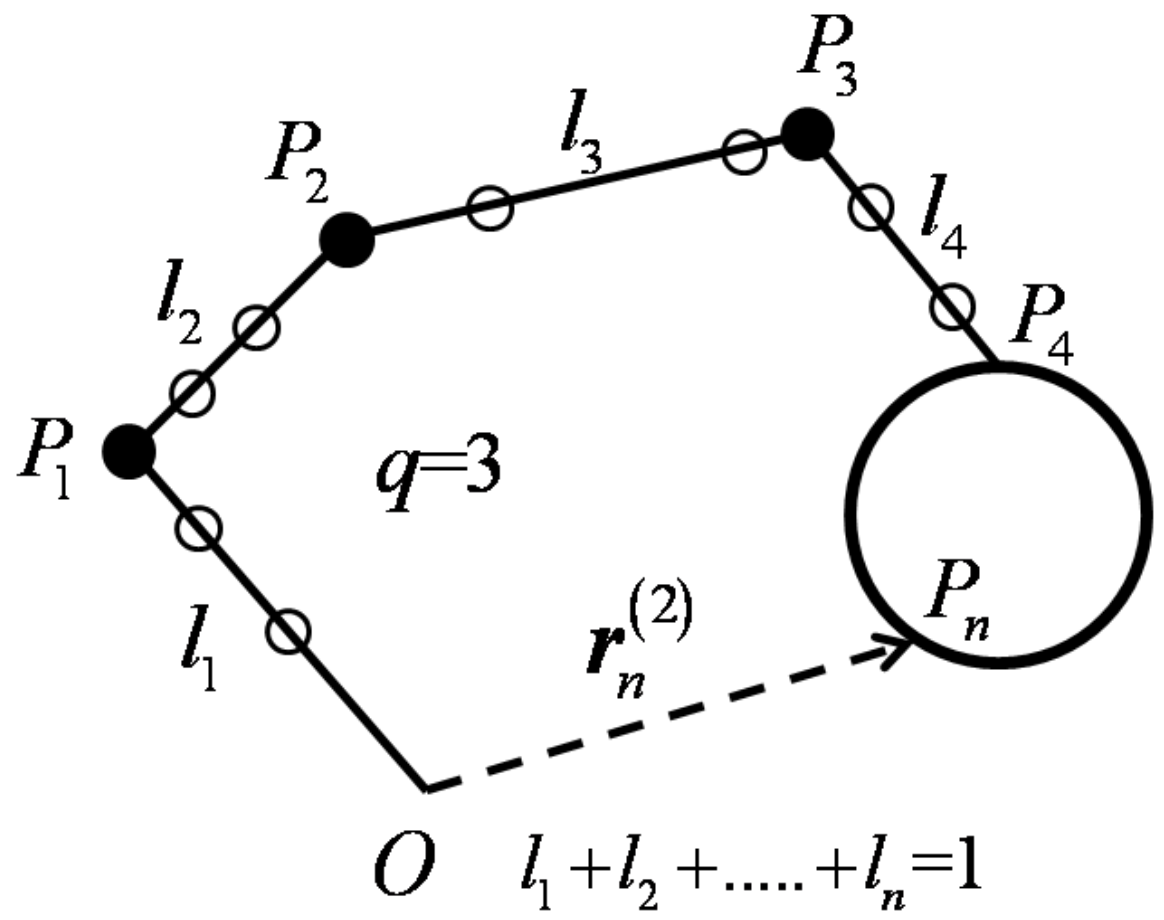

Figure 1: A Pearson-Dirichlet random walk $PD(2,n,q)$ in $\mathbb{R}^2$ for $q=3$: the walk starts at O in a random direction; the length of every step is the sum of $q$ (as indicated by $q-1$ empty circles) i.i.d. exponential random variables; at $P_k\ (k=1,..,n-1)$, a new random direction is taken independently of the previous ones; every step is rescaled so as to make the total travelled length equal to 1. The walk ends at $P_n$ with $\boldsymbol{OP}_n=\boldsymbol{r}_n^{(2)}$. Equivalently, the $n$ step lengths are i.i.d. random variables with a gamma distribution whose shape parameter is $q$.